\documentclass[prxquantum,twocolumn,english,superscriptaddress]{revtex4-2}
\usepackage[utf8]{inputenc}

\usepackage{graphicx}
\usepackage[tight]{units} 
\usepackage{textcomp} 
\usepackage{gensymb} 
\usepackage{dsfont}
\usepackage{amsmath,amsfonts, amssymb}
\usepackage{bm}
\usepackage{hyperref}
\usepackage{xcolor}
\usepackage{braket}

\newcommand{\IQOQI}{\affiliation{Institut f\"ur Quantenoptik  und Quanteninformation, \"Osterreichische Akademie der Wissenschaften, Technikerstra{\ss}e 21a, 6020 Innsbruck, Austria}}
\newcommand{\UIBK}{\affiliation{Institut f\"ur Experimentalphysik, Universit\"at Innsbruck, Technikerstra{\ss}e 25, 6020 Innsbruck, Austria}}
\newcommand{\AQT}{\affiliation{Alpine Quantum Technologies GmbH (AQT), Technikerstra{\ss}e 17/1, 6020 Innsbruck, Austria}}

\begin{document}

\title{Controlling two-dimensional Coulomb crystals of more than 100 ions in a monolithic radio-frequency trap}

\author{Dominik Kiesenhofer}
\email{The first two coauthors contributed equally.}
\UIBK
\IQOQI

\author{Helene Hainzer}
\email{The first two coauthors contributed equally.}
\UIBK
\IQOQI

\author{Artem Zhdanov}
\UIBK

\author{Philip C. Holz}
\AQT

\author{Matthias Bock}
\UIBK
\IQOQI

\author{Tuomas Ollikainen}
\UIBK
\IQOQI

\author{Christian F. Roos}
\email{christian.roos@uibk.ac.at}
\UIBK
\IQOQI
\date{\today}

\begin{abstract}

Linear strings of trapped atomic ions held in radio-frequency (rf) traps constitute one of the leading platforms for quantum simulation experiments, allowing for the investigation of interacting quantum matter. However, linear ion strings have drawbacks, such as the difficulty to scale beyond $\sim 50$ particles as well as the inability to naturally implement spin models with more than one spatial dimension.
Here, we present experiments with planar Coulomb crystals of about 100 $^{40}$Ca$^+$ ions in a novel monolithic rf trap, laying the groundwork for quantum simulations of two-dimensional spin models with single-particle control. We characterize the trapping potential by analysis of crystal images and compare the observed crystal configurations with numerical simulations. We further demonstrate stable confinement of large crystals, free of structural configuration changes, and find that rf heating of the crystal is not an obstacle for future quantum simulation experiments. 
Finally, we prepare the out-of-plane motional modes of planar crystals consisting of up to 105 ions close to their ground state by electromagnetically-induced transparency cooling, an important prerequisite for implementing long-range entangling interactions.
\end{abstract}

\maketitle

\section{Introduction}

Models of interacting quantum many-body systems are often easy to write down but hard to study by analytical or numerical techniques. Quantum simulation embodies the concept that experiments with engineered quantum systems might provide novel insights into quantum many-body physics, making it an important application of noisy intermediate-scale quantum processors \cite{Preskill:2018}. To this end, the experimental system can be directly engineered for a realization of the desired interactions, or sequences of quantum gates can be used for approximating the dynamics in a controlled way \cite{Buluta:2009}, referred to as analog and digital quantum simulation, respectively. Furthermore, entangling interactions may be used as a resource for variational methods \cite{McClean:2016}.

Various experimental platforms have been suggested to realize a quantum simulator---neutral atoms in optical lattices \cite{Gross:2017}, trapped ions \cite{Monroe:2021}, Rydberg atom arrays \cite{Browaeys2020}, superconducting circuits \cite{Arute:2020,Morvan:2022} and photonic waveguide arrays \cite{Aspuru2012, Wang2020photonic}, to name a few. Trapped atomic ions are among the most precisely controllable quantum systems available for applications such as quantum computation \cite{Figgatt2019, Postler2022}, metrology and sensing \cite{Brewer2019, Marciniak2022, Hainzer2022} as well as simulation \cite{Blatt2012, Monroe2021}. In particular, in analog and variational quantum simulation experiments, control over large ion crystals has been proven possible, allowing for the study of quantum many-body physics phenomena with several tens of particles \cite{Zhang2017, Kokail2019, Joshi2022, Li:2023}. 
Scaling up trapped-ion experiments to a larger number of individually controllable particles is challenging, though. Rf traps such as linear Paul traps have enabled complete quantum control of up to 50 ions in linear chains \cite{Kranzl2022}. In these experiments, entangling interactions are mediated across the system by lasers coupling to the collective motional modes of the ion crystal while coherent single-particle operations are achieved by addressing individual ions with tightly focused laser beams. Keeping a long one-dimensional (1d) crystal linear, however, requires extremely anisotropic trapping potentials \cite{Schiffer:1993,Dubin:1993}, eventually limiting the number of ions that can be controlled with high fidelity: Problems in laser cooling arise from the fact that the ions are only weakly confined in the axial direction of the string, resulting in high heating rates of the axial motional modes. In addition, the spatial extent of long linear ion chains leads to difficulties in laser-addressing the outer ions.

One obvious approach to scaling up trapped-ion architectures is to make use of a second spatial dimension. While Penning traps offer the possibility to work with hundreds of ions in a planar configuration \cite{Bohnet2016}, single-ion control has not yet been implemented in the laboratory as these crystals rotate at rates of tens to hundreds of kHz about the out-of-plane axis \cite{Polloreno2022}. Surface trap arrays, where each rf microtrap confines a single ion, would, in principle, allow for ions to be trapped in nearly arbitrary two-dimensional (2d) geometries but so far this has only been achieved in small systems of several ions \cite{Sterling2014, Mielenz2016, Holz2020}. Here, substantial technical challenges need to be overcome: Surface traps suffer from low trap depths and are thus prone to ion loss. 
Furthermore, for the realization of entangling interactions mediated by the ions' motion, the ions need to be brought sufficiently close together (on the order of tens of micrometers) to produce an interaction on timescales much faster than the ions' coherence time. This is only possible by bringing the ions close to the trap electrodes and thus subjecting them to strong surface electric field noise, leading to high heating rates that limit the motional coherence. Lastly, free-space optical access is challenging for large chip traps, creating an additional technological overhead for optics integration. 

Alternatively, it is possible to confine ions in a planar configuration in the potential of a  single rf trap, which is an approach recently taken by several groups around the world \cite{Yoshimura2015, Wang2020, Ivory:2020,Xie2021} and also the one taken for the experiment described in this article.
While this approach allows for harnessing the many benefits that linear ion strings provide, such as tunable-range entangling interactions, reaching from all-to-all to nearest-neighbor couplings, spin models can be implemented in 2d on top of that. On the downside, there are difficulties to be overcome: Most ions in a planar crystal experience rf-driven micromotion, as they are inevitably displaced from the rf null. Any laser used to manipulate these ions will appear to be phase-modulated at the trap drive frequency. However, adverse effects on the laser-ion interactions, which form the basis of laser cooling and coherent operations, can be suppressed to first order by applying laser beams only from directions perpendicular to the micromotion. This potentially opens up another challenge: the necessity to develop novel ion trap geometries in order to gain optical access from relevant directions. 
With such a trap design, promising results have been presented recently, such as electromagnetically-induced transparency (EIT) cooling of the out-of plane motional modes of a 2d crystal consisting of 12 $^{171}$Yb$^+$ ions \cite{Qiao2021} and frustrated quantum magnetism with 10 ions \cite{Qiao2022}. Yet, it remains to be shown that these results can be extended to experiments with considerably larger crystals, which face additional challenges: rf heating of the ions could compromise entangling interactions mediated by ground-state cooled collective modes of motion. Furthermore, planar ion crystals can display various distinct lattice configurations. Transitions between these crystal configurations can be initiated, e.g. by background gas collisions or by rf heating \cite{Block2000, Bluemel1989}. The number of such configuration changes needs to be minimized as single-ion quantum state detection and control rely on stable ion positions.

In this article we present an ion-trap apparatus, designed for quantum simulation experiments with individually-controllable particles in a 2d configuration. We demonstrate quantum control over planar ion crystals of up to 105 ions and provide experimental evidence that crystal lattice configuration changes and rf heating are no obstacles for quantum simulation experiments with these crystals. The paper is structured as follows: In Sec.~\ref{sec:apparatus}, we introduce our novel monolithic linear Paul trap, and describe our approach to loading, cooling, quantum-state manipulation, and readout of planar $^{40}$Ca$^+$ ion crystals. In Sec.~\ref{sec:characterization}, we analyze the planarity and orientation of the ion crystals, characterize the trapping potential, investigate the stability of the lattice configurations, measure the heating dynamics of the ion crystals, and, finally, demonstrate EIT cooling of the out-of-plane modes of a small and a large 2d ion crystal. Our findings are concluded in Sec.~\ref{sec:conclusion}.

\section{Apparatus}
\label{sec:apparatus}

In a standard linear Paul trap, consisting of two rf blades, two dc blades, and two endcap electrodes, planar crystals can be trapped either in the plane spanned by the two radial directions or in a plane spanned by one radial and the axial direction, as illustrated in Figs.~\ref{fig:intro}(a) and \ref{fig:intro}(b), respectively. The necessary condition for the formation of a planar crystal is given by  $\omega_{\textrm{s}}/\omega_{\textrm{w}} >1.23 N^{1/4}$ \cite{Dubin1993}, where $\omega_{\textrm{s}}$ is the secular frequency in the direction of strong confinement, in which the crystal is squeezed flat, $\omega_{\textrm{w}}$ is the frequency in the directions of weak confinement, in which the crystal is extended, and $N$ is the number of ions. The required asymmetry of the trap frequencies can be achieved by introducing dc voltages on the dc and endcap electrodes to squeeze the crystal into the desired plane.

In terms of the rf power requirements, neither of the two crystal orientations provides an advantage compared to the other one (see App.~\ref{sec:calculations_appendix}). However, it is beneficial to choose the orientation shown in Fig.~\ref{fig:intro}(b) because rf-driven micromotion occurs only along a single direction. In this case an entire plane is available from which the ions can be addressed with laser beams perpendicular to the micromotion, therefore eliminating the adverse effects on laser-ion interactions to the first order.

We therefore choose the crystal orientation illustrated in Fig.~\ref{fig:intro}(b), and modify the standard linear Paul trap geometry as shown in Fig.~\ref{fig:intro}(c), to gain optical access from relevant directions for laser cooling and manipulation, and in particular perpendicular to the crystal plane for ion imaging and laser-addressing of individual ions. The price to pay for having a micromotion-free optical access in an entire plane is a slightly increased sensitivity of the trap potential anisotropy with respect to rf-voltage fluctuations (see App.~\ref{sec:calculations_appendix}), which can be mitigated by a state-of-the-art active stabilization of the rf voltage.

In the following, we provide a detailed description of our linear Paul trap and overall apparatus for experiments with planar crystals of $^{40}$Ca$^+$ ions.

\subsection{A novel monolithic linear Paul trap}

\begin{figure}[tbp]
    \centering
    \includegraphics[width=.95\textwidth]{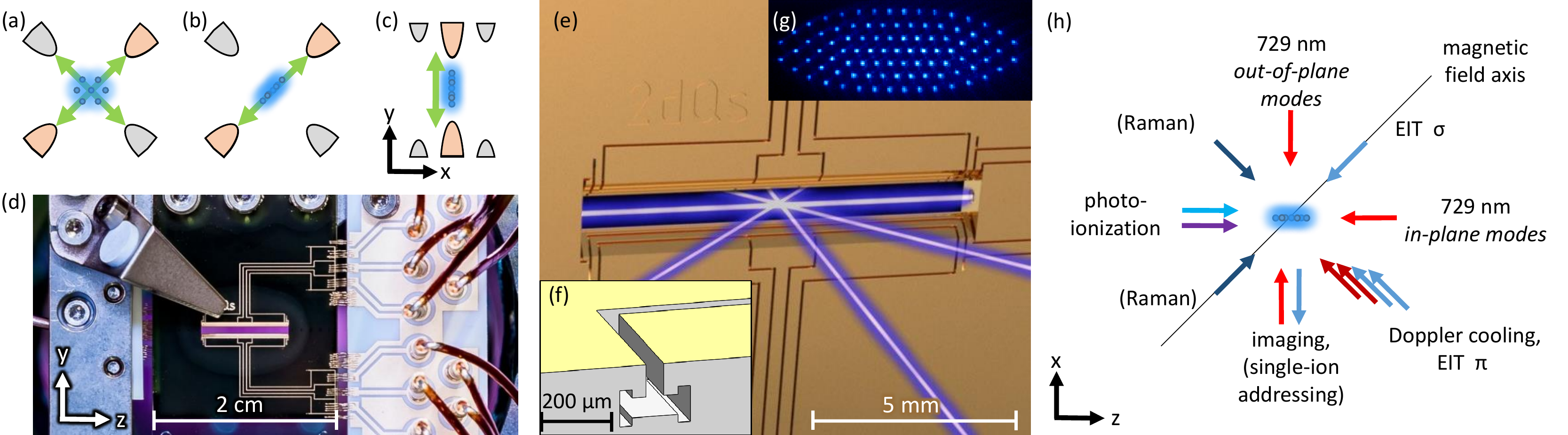}
    \caption[format=plain,justification=justified,singlelinecheck=false]{
    (a,b) Schematic radial cross section of a linear Paul trap. The orange (grey) shapes symbolize rf (dc) electrodes. The green arrows denote the directions of micromotion experienced by the ions in the crystal (blue). In (a), the planar crystal is oriented in the plane spanned by the radial directions of the trap, and in (b), by one radial and the axial direction. (c) Schematic radial cross section  of a three-layer trap: This geometry allows for optical access perpendicular to micromotion. (d) Photograph of the ion trap inside the vacuum chamber: The trap chip is seen in the center of the picture. Ions are loaded from a calcium target, attached to a funnel aperture (top left). Two printed-circuit-boards made of alumina (front right and rear left of the trap chip) provide interfaces between gold wirebonds to the trap electrodes and cables to the vacuum feedthroughs. (e) 3D rendering of the central region of the ion trap (rear side with rf electrode leads) and laser beams in the horizontal plane. (f) Schematic illustration of the trench geometry, used to electrically isolate electrodes. The electrode separation on the surface of the trap chip is 50\,\textmu m. (g) A Coulomb crystal consisting of 91 $^{40}$Ca$^+$ ions stored in the center of the trap. (h) Top view of the laser beam and magnetic field geometry in the horizontal plane. Parentheses indicate beams that have not yet been implemented. Not shown are two laser beams which do not lie within the horizontal plane: A 729\,nm beam for spectroscopy of in- and out-of-plane motional modes which has a 45$^\circ$ angle with the horizontal plane, and a second cooling beam at a $\sim 20^\circ$ angle with the horizontal plane coupling also to the vertical direction.}
    \label{fig:intro}
\end{figure}

Effectively, our electrode geometry, shown in Fig.~\ref{fig:intro}(c), realizes a three-layer trap \cite{Deslauriers:2004}: It differs from a standard linear Paul trap in that the dc blades are split and projected to the top and bottom. The distance between the rf electrode and the trap center is 400~\textmu m. The dc electrodes are recessed with respect to the rf electrode in order to allow for additional optical access at 45$^\circ$ angles from top and bottom. Furthermore, these electrodes are segmented as shown in Fig.~\ref{fig:intro}(e). The outer segments represent the endcap electrodes and allow for axial confinement. Additional ground electrodes are introduced between the rf and dc electrodes. Their purpose is to reduce rf field components in the axial trap direction introduced by the gaps between dc segments, which can lead to unwanted micromotion of the ions along that direction. More details can be found in App.~\ref{sec:trap_appendix}.

While machined and hand-assembled linear Paul traps have led to state-of-the-art experiments with 1d ion crystals, small alignment imperfections could lead to difficulties in working with large 2d ion crystals. A misalignment of the trap electrodes with respect to each other could, on the one hand, enhance nonlinear resonances in the trapping potential which can lead to the ejection of ions from the trap \cite{vBusch1961, Alheit1996}, an effect which increases in severity the further an ion is displaced from the rf null \cite{Blakestad2010}. On the other hand, a misalignment could lead to increased micromotion in directions which cannot be effectively compensated. 
Therefore, we designed a monolithic ion trap allowing for precise electrode arrangements without the need for hand assembly. The trap is shown in Fig.~\ref{fig:intro}(d). Figure~\ref{fig:intro}(e) displays a rendering of the central trap region with laser beams representing the directions used for optical access in the $xz$ (horizontal) plane. Holes in the trap chip enable laser access from the axial direction. Planar ion crystals are trapped in the $yz$ plane, an example comprised of 91 ions is shown in Fig.~\ref{fig:intro}(g).

The trap was microfabricated \footnote{Translume Inc., Ann Arbor, MI 48108, USA} via selective laser-induced etching \cite{Bado2015, Noel2019, Ragg2019, Decaroli2021}. In the first stage of the manufacturing process the design is written into a fused silica wafer with a pulsed laser, locally changing the properties of the glass. The illuminated material is etched in a bath of hydrofluoric acid in the second stage. Using this approach it is possible to realize electrode structures with sub-micrometer precision, and, in particular, to create trenches that extend underneath the surface of the chip \cite{Bado2015, Araneda2020}. After metallization (sputtering of 30\,nm titanium and 3\,\textmu m gold) the trenches with undercuts, shown in Fig.~\ref{fig:intro}(f), keep the electrodes electrically isolated. 

The ion trap is operated at rf voltages of $\sim$ 1 kV peak-to-peak and with a frequency of $\Omega_{\textrm{rf}} = 2 \pi \times 43.22$ MHz in order to ensure a Mathieu stability parameter of $q$ \small 
$\lesssim$ \normalsize $ 0.1$ and, thus, a low level of micromotion. In the experiments presented in this article the crystals are trapped in a potential with secular oscillation frequencies of $\omega_{\textrm{s}}/(2\pi) \approx 2.2$ MHz in the strongly confining direction and a few hundred kHz in the two weakly confining directions. The rf power is actively stabilized in order to ensure stability of the confinement on both long and short time scales. More details on control and stability of the trap rf can be found in App.~\ref{sec:stab_appendix}.

Since the principal axes of the trap potential are well aligned with the trap's geometrical axes,  $x$, $y$, and $z$, the secular frequencies are henceforth denoted as $\omega_x$, $\omega_y$, and $\omega_z$, respectively, where $x$ is the direction perpendicular to the ion crystal plane.

\onecolumngrid\
\twocolumngrid\

\hfill

\subsection{Laser beam and magnetic field geometry}

Figure~\ref{fig:intro}(h) provides a schematic overview of the geometry of the laser beams and magnetic field in our apparatus. The quantization axis is defined via permanent magnets, creating a magnetic field that is parallel to the $xz$ plane and has an angle of 45$^\circ$ with respect to the crystal plane. Additionally, a pair of coils in anti-Helmholtz configuration along the same direction is used to compensate unwanted magnetic field gradients.

We use eight directions for optical access in the $xz$ plane, which is perpendicular to the direction of micromotion, and thus allows for micromotion-free laser-ion interaction. Along $x$, an objective lens, with a numerical aperture of 0.289, is employed to collect fluorescence light at 397\,nm from the ions for imaging and state readout. The light is guided and focused onto the chip of an electron-multiplying CCD (EMCCD) camera and a photomultiplier tube (PMT). Two photoionization beams are focused through the axial holes of the trap (along $z$). The ions are Doppler-cooled from a direction orthogonal to the quantization axis, from which we can couple to all in- and out-of-plane motional modes with the exception of the center-of-mass mode along $y$. However, we find that ion crystals can still be cooled efficiently in all three directions, with the exception of small crystals of several ions, or single ions, for which we employ a different cooling beam that has an overlap with the $y$ direction (20$^\circ$ angle with horizontal plane, not shown in Fig.~\ref{fig:intro}(h)). Ground-state preparation of the out-of-plane motional modes by means of EIT cooling is carried out using two beams: one circularly polarized propagating along the quantization axis, and one linearly polarized propagating perpendicular to it. Coherent operations on the optical qubit transition at 729\,nm are implemented with beams along the $x$ and $z$ direction. These beams can also be used to couple to the out-of-plane motional modes, or the in-plane modes, respectively. Moreover, a 729\,nm beam at an angle of 45$^\circ$ with the horizontal plane (not shown) couples to both in- and out-of-plane motional modes.

The beams labeled in parentheses in Fig.~\ref{fig:intro}(h) are not yet implemented in our setup: In the future we aim to realize entangling interactions by coupling to the out-of-plane motional modes with bichromatic Raman laser beams at 396\,nm. These will propagate along and perpendicular to the quantization axis, respectively, opposite to the EIT cooling beams. Additionally, we plan to use the objective lens to tightly focus laser beams at 729\,nm for addressing of individual ions.

\subsection{Loading and cooling}

Ions are loaded via laser ablation using a 515\,nm pulsed laser. The power of the ablation pulses can be precisely controlled via a motorized waveplate in combination with a polarizing beam splitter and is set low enough such that mostly neutral atoms are ablated from the calcium target. For isotope-selective photoionization of $^{40}$Ca we employ a two-stage process: A 422\,nm laser drives a cycling transition to a highly excited state from which the valence electron is detached by laser light with a wavelength of 375\,nm. In the trap center, the atomic beam and the photoionization beams cross at an angle of 45$^\circ$, which allows for the exploitation of the Doppler shift to select slower atoms from the velocity distribution. 

Deterministic loading of single ions and large crystals with a desired ion number is typically realized on a timescale of less than 1 minute by combining control over ablation laser power (to add ions to the crystal) and over the trapping potential depth (to remove ions from the crystal). 

Doppler cooling is performed on the $4\textrm{S}_{1/2} \leftrightarrow 4\textrm{P}_{1/2}$ dipole transition at 397\,nm, with two repumper lasers at 854\,nm and 866\,nm which remove population from the long-lived D states. Further cooling of the ions to the motional ground state is possible via either sideband or EIT cooling. For EIT cooling, we employ 397\,nm light that is blue-detuned by 110\,MHz from the $4\textrm{S}_{1/2} \leftrightarrow 4\textrm{P}_{1/2}$ transition, as discussed in more detail in Sec.~\ref{sec:groundstatecooling}.

\subsection{Qubit manipulation}

Qubits can be encoded in the two Zeeman states of the $4\textrm{S}_{1/2}$ ground-state manifold (ground-state qubit) or in one of the two ground states in combination with one of the metastable $3\textrm{D}_{5/2}$ Zeeman states (optical qubit). We couple the ground states coherently via a magnetic rf field, oscillating at approximately 11\,MHz, which is generated via a coil placed outside of vacuum. We enclose our apparatus in a magnetic-field shielding and obtain a coherence time of 94(14)~ms for the ground-state qubit via Ramsey experiments. A standard spin-echo pulse sequence extends the ground-state coherence time to 251(27)~ms.

The optical qubit is manipulated with a frequency-stable ($\sim 1$~Hz linewidth) laser at 729\,nm to realize coherent operations on the electronic $4\textrm{S}_{1/2} \leftrightarrow 3\textrm{D}_{5/2}$ quadrupole transition.
In order to distinguish the two qubit states we employ the electron-shelving technique: Ions are excited on the $4\textrm{S}_{1/2} \leftrightarrow 4\textrm{P}_{1/2}$ transition and their fluorescence is detected via the PMT and imaged onto the EMCCD camera. An ion appears bright or dark, depending on whether the electron is projected onto the $4\textrm{S}_{1/2}$ or $3\textrm{D}_{5/2}$ state of the optical qubit.
We map the ground-state qubit to the optical qubit using 729\,nm light and employ the same read-out technique.

\subsection{Quantum state analysis}

Data analysis for a system comprised of more than one ion relies on ion-specific readout of bright and dark states from camera images. Here, a high-fidelity distinction between the qubit states, even for short detection times of a few milliseconds, is crucial. In the case of 2d crystals, this becomes even more challenging than for linear ion chains, as the ions can suffer from higher levels of light scattering crosstalk due to the increased number of neighboring ions.

Our algorithm for quantum state analysis will be presented in detail in a forthcoming publication. Here, we provide just a short summary:
The general idea is to first take a set of reference images, in the beginning of the experiment, where the ions are randomly bright and dark. Next, we employ statistical tools to obtain projection matrices which can be multiplied to an image yielding the pseudocounts for each ion, being free of crosstalk counts from neighbouring ions.

\section{System Characterization}
\label{sec:characterization}

Our apparatus is capable of trapping stable crystals of up to $N\approx100$ ions. We observe crystal lifetimes of several hours or even days, limited only by the occurrence of a ``dark ion". This can be either a different atomic species, other than $^{40}$Ca$^+$, which is caught in the trapping potential, or a molecular ion, formed in a chemical reaction between $^{40}$Ca$^+$ and the residual background gas. In order to quantify the achievable control over ions in our setup and assess the suitability of our platform for future quantum simulation experiments, characterization measurements are performed. We investigate the planarity, orientation, heating dynamics, crystal lattice stability, and ground-state cooling of large 2d ion crystals, as well as the trap potential anharmonicity by analysis of crystal images.

\begin{figure}[tbb]
    \centering
    \includegraphics[width=0.45\textwidth]{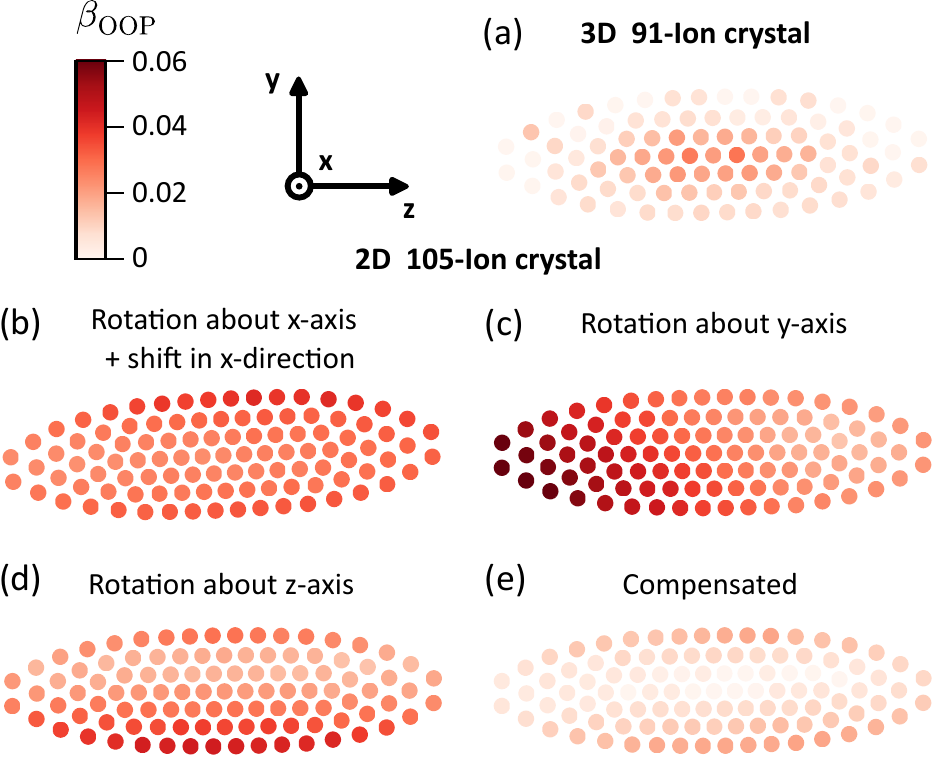}
    \caption[format=plain,justification=justified,singlelinecheck=false]{Measurement of the micromotion modulation index in out-of-plane direction. (a) A 91-ion crystal is employed to investigate planarity. In this example, ions in the center of the crystal display a higher modulation index as they are pushed out to the third dimension. (b--d) The images show examples of planar 105-ion crystals displaying various undesirable scenarios where shifts along and rotations about arbitrary axes are present, leading to increased micromotion modulation indices. The labels indicate the main contribution to the crystal misalignment. (e) By applying pre-calculated sets of dc voltages, we are able to compensate for shifts and rotations and align the ion crystal with the trap center and principal axes such that the micromotion modulation indices are minimized for all ions.}
    \label{fig:planarity}
\end{figure}

\subsection{Planarity and crystal orientation}

In order to ensure the planarity and optimum alignment of an ion crystal in the trap center, we probe the sideband-resolved 4S$_{1/2}\leftrightarrow$ 3D$_{5/2}$ transition. We determine the out-of-plane micromotion modulation index $\beta_{\textrm{OOP}}$ of the individual ions by measuring the micromotional sideband Rabi frequency $\Omega_{\textrm{sb}}$ and the respective carrier Rabi frequency $\Omega_{\textrm{carr}}$ with a laser beam whose incidence is normal to the plane in which we confine the crystal.  In the limit of small modulation, we have $\beta_{\textrm{OOP}} \approx 2 \, \Omega_{\textrm{sb}}/\Omega_{\textrm{carr}}$.
In Fig.~\ref{fig:planarity}, the measured modulation indices of individual ions in a 91-ion and 105-ion crystal are shown. When the crystal is in an optimum position in the center of the trap we expect the modulation indices to be minimized. However, several imperfections can arise, such as the ions not being confined in a single plane, or the crystal plane being misaligned with respect to the trap center or tilted with respect to the principal axes. In Fig.~\ref{fig:planarity}(a), we investigate the planarity of a 91-ion crystal. In this example, the confinement in the $x$ (out-of-plane) direction is too low and, consequently,  the ions in the center of the crystal display a higher micromotion modulation index as they are pushed out of the crystal plane. This can be resolved by increasing the voltages applied to the middle segments of the ion trap dc electrodes.

Figures~\ref{fig:planarity}(b--e) display the micromotion modulation indices for 105-ion planar crystals. Figures~\ref{fig:planarity}(b--d) show examples where the crystals are not optimally positioned and oriented in the trap center. Here, we observe tilts and shifts, reflected in a higher level of micromotion for certain ions. Such tilts or shifts can be a consequence of asymmetries in the trap geometry as well as stray fields, e.g., from dust particles. However, we can compensate for these by applying asymmetric voltages on the outer segmented electrodes of the trap, to tilt and translate the crystal in all three dimensions and move it into the optimum position in the trap center. In this case---combined with applying enough voltage to the middle segments to ensure planar crystals---we end up with a well-compensated crystal shown in Fig.~\ref{fig:planarity}(e), where $\beta_{\textrm{OOP}} < 0.02$ for all ions. The slightly increased modulation indices of the outer ions likely stem from a small misalignment of either the laser beam or the ion crystal, such that the beam is not perfectly perpendicular to the crystal plane. However, these effects are small, reflected in the overall low modulation indices, and do not restrict the planned experiments. For the optimally-placed 2d crystals consisiting of 91 and 105 ions, we measure the oscillation frequencies $\omega_{x,y,z} = 2\pi\times(2196, 680, 343)$\,kHz and $\omega_{x,y,z} = 2\pi\times(2188,528,248)$\,kHz, respectively.

\subsection{Crystal shape}
\label{sec:shape}
A planar crystal trapped in a harmonic potential with anisotropy takes on an elliptical shape. The latter is quantified by the crystal's aspect ratio $\zeta=a_2/a_1$, defined as the length ratio of the semi-minor axis $a_2$ and the semi-major axis $a_1$. When carrying out numerical simulations of ion positions in a given harmonic potential, knowledge of $\zeta$ is helpful for starting the simulation with a good initial guess of the ion positions. However, there is no simple relation linking $\zeta$ to the trapping frequencies. 
We determine $\zeta$ in the following way: We calculate the covariance matrix $C$ of the ion positions $(y_i,z_i)$, which can be precisely measured by recording the ions' fluorescence on a CCD camera \cite{Kaufmann:2012}. Diagonalization of $C$ yields $\zeta =\sqrt{\lambda_2/\lambda_1}$ where $\lambda_1\ge\lambda_2$ are the two eigenvalues of $C$. In a harmonic potential characterized by its trap anisotropy 
\begin{equation}
\xi = \omega_y/ \omega_z,
\end{equation}
one finds $\zeta\le\xi^{-1}$. For sufficiently large ion numbers, the crystal can be modeled as a charged fluid, the shape of which can be calculated from potential theory \cite{Kellogg:1929}. Then, $\zeta$ is related to $\xi$ via 
\begin{equation}
\zeta^2\frac{K-E}{E-\zeta^2K}=\xi^{-2} \label{eq:aspectratio}
\end{equation}
where $K = K(\sqrt{1-\zeta^2})$ and $E = E(\sqrt{1-\zeta^2})$ are complete elliptic functions of the first and second kinds [cf. Eq.~(3.81) of Ref.~\cite{Dubin:1999}]. 

We apply this method for a measurement of the aspect ratio to a 91-ion crystal. From sideband spectroscopy on the 4S$_{1/2} \leftrightarrow$ 3D$_{5/2}$ transition the secular oscillation frequencies are determined to be $\omega_y = 2 \pi \times 742\,$ kHz and $\omega_z = 2 \pi \times 370\,$ kHz, yielding an anisotropy of $\xi^{-1}= 0.499$. From the ion positions we find an aspect ratio of $\zeta = 0.382$, in decent agreement with $\zeta_{th}=0.403$ predicted by Eq.~(\ref{eq:aspectratio}).

\subsection{Characterization of the trapping potential from crystal images}

\begin{figure}[tbp]
    \centering
    \includegraphics[width=.49\textwidth]{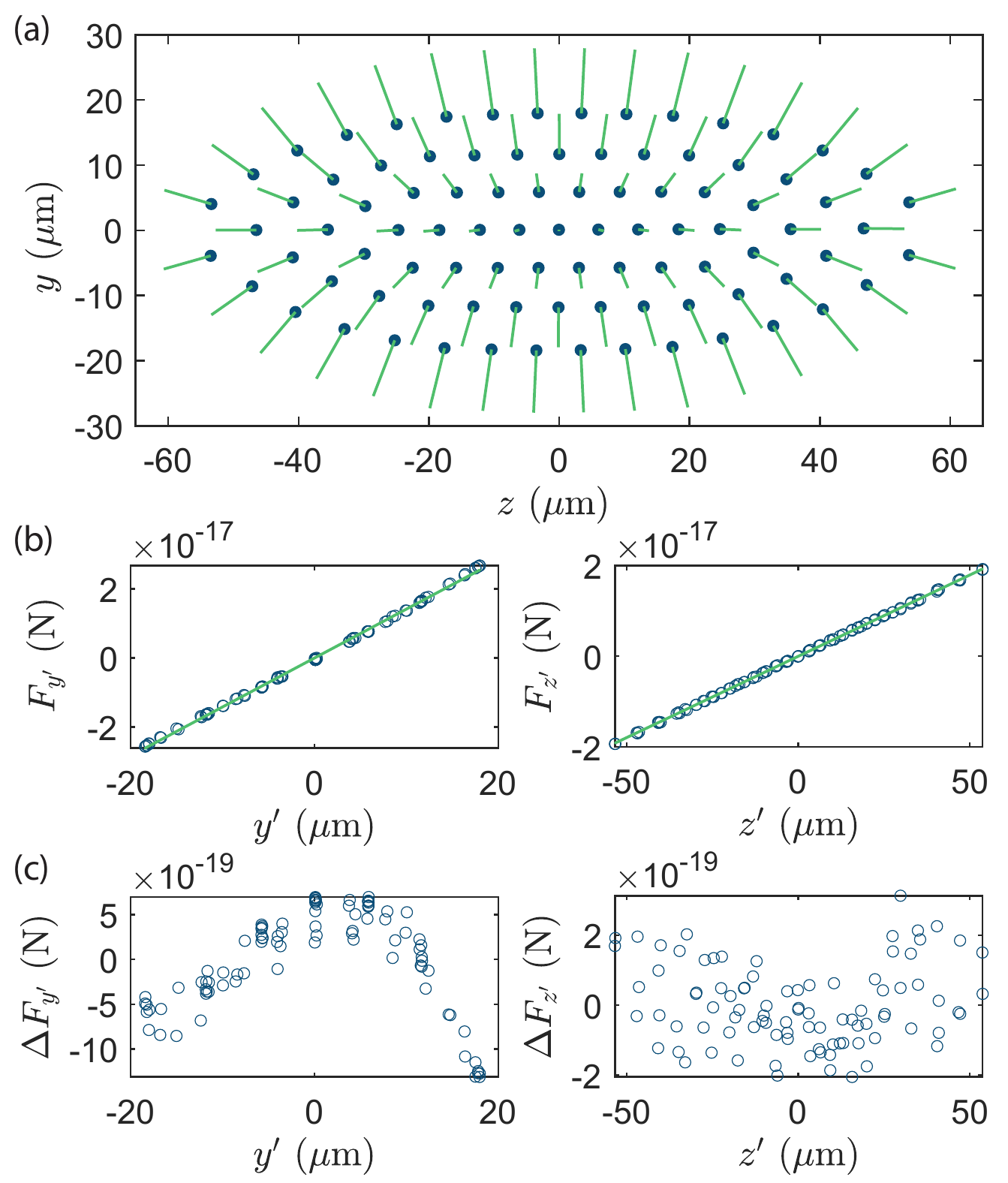}
    \caption[format=plain,justification=justified,singlelinecheck=false]{Characterization of the trapping potential from an image of a 91-ion planar crystal. (a) The blue points indicate the positions of the individual ions, green lines represent the forces exerted on the ions (arbitrary units). (b,c): Force components $F_{y'}$ and $F_{z'}$ and their residuals $\Delta F_{y'}$ and $\Delta F_{z'}$ (with respect to the linear fit) along the two directions $y'$ and $z'$ of the rotated coordinate system.}
    \label{fig:pot_char}
\end{figure}

The analysis of planar crystal images is not limited to determining the anisotropy of the trapping potential via Eq.~(\ref{eq:aspectratio}), but yields much more information such as trapping frequencies and possible anharmonicities of the potential:  The ions‘ equilibrium positions in a crystal are determined by a balance between the confining force of the trap potential and the repelling Coulomb force between the ions. For a planar crystal, we can calculate the Coulomb forces between the ions from their observed positions and, therefore, deduce the force that is exerted by the trapping potential. 
Assuming a harmonic trapping potential
\begin{equation}
V = \frac{1}{2}\begin{pmatrix} z-z_0, & y-y_0 \end{pmatrix} A \begin{pmatrix} z-z_0 \\ y-y_0 \end{pmatrix}
\end{equation}
where $A = \begin{pmatrix} \alpha & \gamma \\ \gamma & \beta \end{pmatrix}$ is a symmetric matrix, $V$ is determined from a fit after calculation of the Coulomb force components in the two directions spanning the crystal plane (vertical ($y$) and horizontal ($z$) on the camera image). Figure~\ref{fig:pot_char}(a) shows the Coulomb force field that we deduce from an image analysis of a 91-ion crystal. The eigenvalues of $A$ are proportional to the squares of the secular oscillation frequencies in the crystal plane. Taking the square root of the ratio of these oscillation frequencies, we find a potential anisotropy of $\xi^{-1} = 0.503$, in good agreement with the result $\xi^{-1}= 0.499$ obtained from sideband spectroscopy in Sec.~\ref{sec:shape}. The principal axes of $V$ are not necessarily aligned with the coordinate system $(z,y)$ of the imaging system. For this reason, we introduce a rotated coordinate system defined by $z^\prime=z\cos\phi+y\sin\phi$ and $y^\prime=-z\sin\phi +y\cos\phi$. The fit of $V$ yields $\phi=0.004$, showing that both coordinate systems nearly coincide for this crystal. In Fig.~\ref{fig:pot_char}(b), we plot the Coulomb force components $F_{y^\prime}(z_i^\prime,y_i^\prime)$, $F_{z^\prime}(z_i^\prime,y_i^\prime)$ along the principal axes; the residuals $\Delta F_{y^\prime}$, $\Delta F_{z^\prime}$ from a linear fit are shown in Fig.~\ref{fig:pot_char}(c). These residuals reveal that the trapping potential is not purely harmonic. However, as the deviations from the linear fit are small, we conclude that anharmonic contributions to the trapping potential are rather minor. The determination of the Coulomb forces in physical units as presented in Fig.~\ref{fig:pot_char} necessitates a calibration of the imaging system which we obtain by matching the value of the secular frequency $\omega_z$ found by crystal image analysis to the one obtained by sideband spectroscopy.

Once the magnification of the imaging system has been precisely calibrated, image analysis of planar ion crystals might enable measuring the dependence of the potential curvatures on changes of particular dc trap voltages much faster than approaches based on sideband spectroscopy.

\subsection{Crystal configurations and stability}

Langevin collisions of the ions with atoms and molecules of the residual background gas can induce a transition to a non-crystalline (melted) phase in which energy can be transferred from the trapping rf field to the ions in the cloud \cite{vanMourik2022, Bluemel1988}. A melted crystal prevents any further quantum measurement. Hence, during an experimental sequence it is essential that the ion crystal maintains its crystalline structure, which we verify with quantitative measurements described in the following subsection.

Furthermore, lower-energy collisions can initiate transitions between distinct configurations of the crystalline structure of a planar ion crystals held at a fixed trap potential. This phenomenon is for example well known for a zig-zag crystal with two mirror-symmetric degenerate configurations at minimum potential energy \cite{Block2000, Fishman2008}. For larger, less elongated 2d ion crystals, we observe an increasing number of distinct crystal configurations. We analyze those by recording image series of the ion crystal as described below. With this tool we are able to detect, minimize, and filter out the occurrence of metastable lattice configurations during experimental data taking, which is critical as individual-ion read out and laser addressing rely on stable ion positions.

\subsubsection{Melting and recrystallization}

\begin{figure}[tbp]
    \centering
    \includegraphics[width=0.47\textwidth]
    {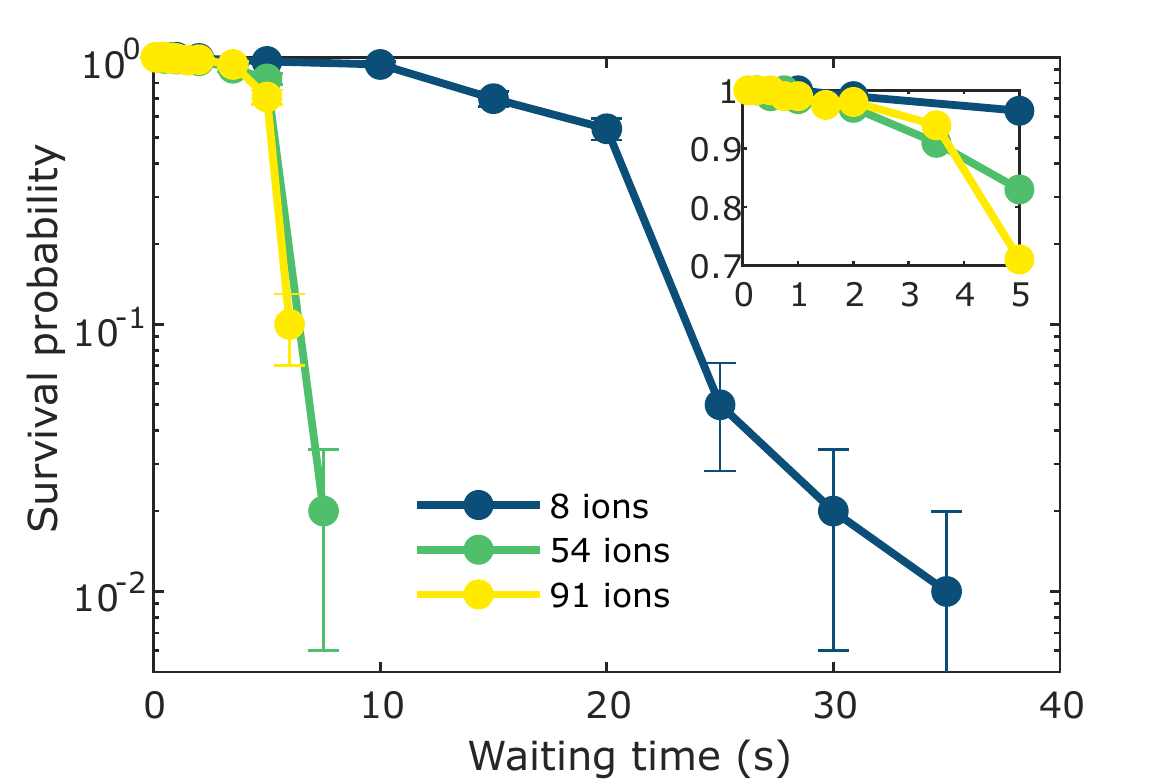}
    \caption[format=plain,justification=justified,singlelinecheck=false]{Melting behavior of planar crystals. Shown here is the probability that a crystal consisting of a certain number of ions has not melted into a cloud after a waiting time without any laser cooling. For each data point 100 experiments were performed. The inset shows the probabilities for short waiting times in more detail.}
    \label{fig:melting}
\end{figure}

\begin{figure*}[tbp]
    \centering
    \includegraphics[width=.9\textwidth]{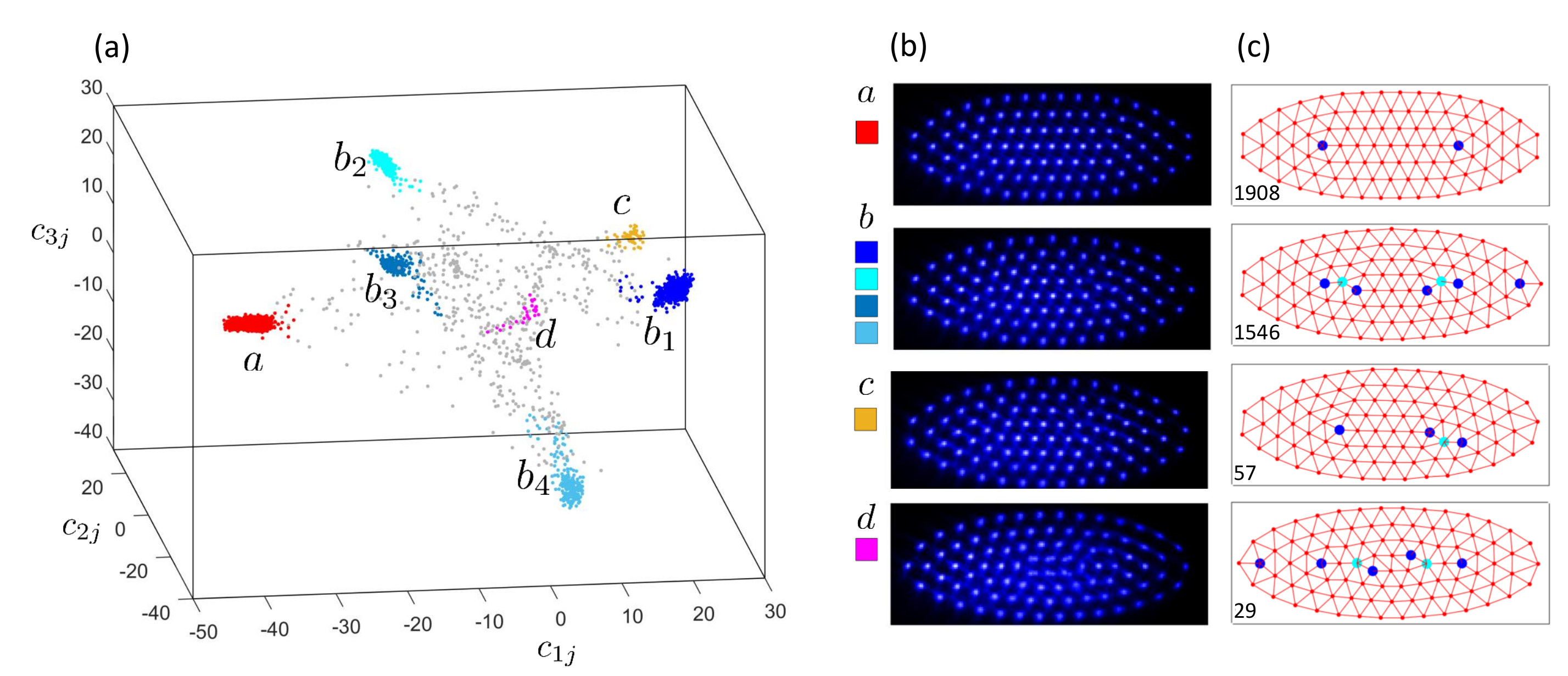}
\caption[format=plain,justification=justified,singlelinecheck=false]{Configuration analysis of a $91$-ion crystal using 4100 camera images. (a) Each image is represented as a point in state space spanned by the projection coefficients $c_{ij}$. Here, only the coordinates corresponding to the three highest eigenvalues are plotted. The clustering algorithm identified seven different clusters ($a$, $b_1,\ldots b_4$, $c$, $d$). Points are colored depending on their cluster label. Grey points are unlabeled and correspond to images where the ions are hot or undergoing a configuration change during the image acquisition time. (b) Crystal configurations representing the cluster centers. Cluster configurations $b_i$ can be converted into each other by sign flips of the picture coordinates. (c) Each configuration can be represented by a planar graph whose vertices represent ions and edges link nearest neighbors. The ions form a triangular lattice with some defects where ions have five or seven neighbours. These defects are plotted as dark blue and light blue vertices, respectively. The number of images found in one of the four configurations is indicated in the lower left corner.}
    \label{fig:cluster_algo}
\end{figure*}

When a melting event occurs, less photons are scattered by the nonlocalized, hot ions during Doppler cooling. The resulting fluorescence count drop on the PMT is used to automatically detect these events during the measurement sequences and trigger immediate recrystallization of the ions. For this purpose, a far red-detuned ($\Delta = 2 \pi \times 330$~MHz) Doppler cooling beam is employed for 100 ms along with the primary Doppler cooling beam. We quantify the time it takes planar ion crystals consisting of 8, 54, and 91 ions to melt into an ion cloud in the absence of any laser cooling in the following way: Cooling beams are turned off for a variable waiting time after which fluorescence counts are recorded with the PMT. These counts allow for a distinction between crystalline and non-crystalline phases. Afterwards, Doppler cooling and recrystallization beams are turned on for 100\,ms to retrieve the crystalline structure of the ions. At the end of the cycle a camera image is taken to ensure that the ions are recrystallized before the next measurement. Figure~\ref{fig:melting} shows the percentage of measurements in which an ion crystal survives in crystalline form as a function of the waiting time. The time it takes for an 8-ion crystal to be melted in 100 percent of cases is $\sim 40$\,s. For larger crystals the melting events occur on a shorter timescale, but still the ions remain in crystalline phase with high probability even after several seconds. Assuming only Langevin collisions as a cause for melting, one would expect a purely exponential behavior as a function of time. We interpret the observed nonexponential decay and the nonlinear dependence on the ion number as evidence that these melting events are not exclusively caused by collisions with residual background gas, but possibly by a combination of collisions followed by rf-induced heating. Regardless of the mechanism behind, the observed time scale of several seconds for larger crystals is long compared to the one of the envisaged experimental sequences.

A further experiment is performed to estimate the time it takes to recrystallize an ion cloud after the occurrence of a melting event. Here, cooling lasers are switched off for a waiting time of 1\,s. This represents an upper limit for planned experiments and, thus, mimics a realistic time scale for the free evolution time of the ion cloud before the recrystallization pulse is applied.  After 1\,s  the ions' fluorescence is measured by the PMT to determine whether the ion crystal has melted. When a melting event is detected, a recrystallization pulse of variable length is applied before an image of the ions is taken in order to check if the recrystallization was successful. In a 91-ion crystal we observe recrystallization after 70\,ms in 101 out of 102 observed melting events. Hence, a pulse duration of 100 ms is sufficient to recrystallize a cloud after a melting event is detected within a measurement sequence.

\subsubsection{Crystal configuration analysis}

In order to detect distinct crystal lattice configurations we record a series of $N_\textrm{P}$ images taken approximately half a second apart, with $N_\textrm{P}\approx 10^4$. Every image $j$ is represented as a column vector $\mathbf{x}_j$ of pixel brightness values and stored in a matrix $M=[\mathbf{x}_1 \mathbf{x}_2\ldots \mathbf{x}_{N_P}]$ where columns and rows correspond to individual images and pixels, respectively. A dimensionality reduction is achieved by means of a principal component analysis of $M$, yielding a number of ``eigenpictures" $\mathbf{y}_i$ that correspond to the eigenvectors of the largest eigenvalues of the covariance matrix $Q=MM^T$.

These eigenpictures provide a basis that enable a faithful representation of any image $\mathbf{x}_j\approx\sum_{i=1}^n c_{ij}\mathbf{y}_i$, where $c_{ij}=\mathbf{y}_i^T\mathbf{x}_j$, using only a small number $n$ of eigenpictures. In this way, every picture can be represented by a set of $n$ numbers. We empirically find that $n\leq 8$ is sufficient for distinguishing different crystal configurations in our experiments. In a final step, the clustering algorithm DBSCAN \cite{Ester1996} identifies clusters in the reduced state space that correspond to different crystal configurations. This routine assigns each image of the recorded series to a crystal configuration whose occurrence over time is analyzed.

Figure~\ref{fig:cluster_algo} illustrates the procedure for a set of $N_P=4100$ pictures of a $91$-ion crystal. In Fig.~\ref{fig:cluster_algo}(a), each image is represented by a point defined by its three most important coefficients $c_{ij}$, $i\in\{1,2,3\}$. Colors are used to indicate the cluster to which each image has been assigned. The clustering algorithm identifies seven different crystal configurations shown in Fig.~\ref{fig:cluster_algo}(b). These configurations can be represented by the planar graphs shown in Fig.~\ref{fig:cluster_algo}(c): The ions form triangular lattices with a certain number of defects that account for the fact that the distance between neighboring ions increases the further the ions are away from the crystal center. For each configuration, we determine the probability $p_i$ of observing it by counting the number of images $N_i$ assigned to its cluster and normalizing it, $p_i=N_i/N_P$. In this data set, the main configuration is found with $\max(p_i)<0.5$, which might seem problematic for quantum simulation experiments for which the crystal always has to be prepared in the same configuration. It is, however, possible to substantially increase this number by fine-tuning the trapping potential such that the energy gap between the ground-state structural configuration and metastable ones increases. To this end, we vary the ratio of oscillation frequencies, $\xi = \omega_{y}/\omega_{z}$, along the two weakly confining directions with the goal of maximizing $p_i$ for the main configuration. Figure~\ref{fig:sim_annealing}(a)  displays measurements of $p$ as a function of the trap anisotropy, showing that for $\xi=1.987$ configuration $a$ is found in more than 99\% of the images. Note that in contrast to the case of barely two-dimensional zig-zag ion crystals where there are two degenerate ground-state configurations, there is only a single ground state because of the mirror symmetries of this crystal configuration.

\subsubsection{Crystal configuration simulations}

\begin{figure}[tbp]
    \centering
    \includegraphics[width=.5\textwidth]{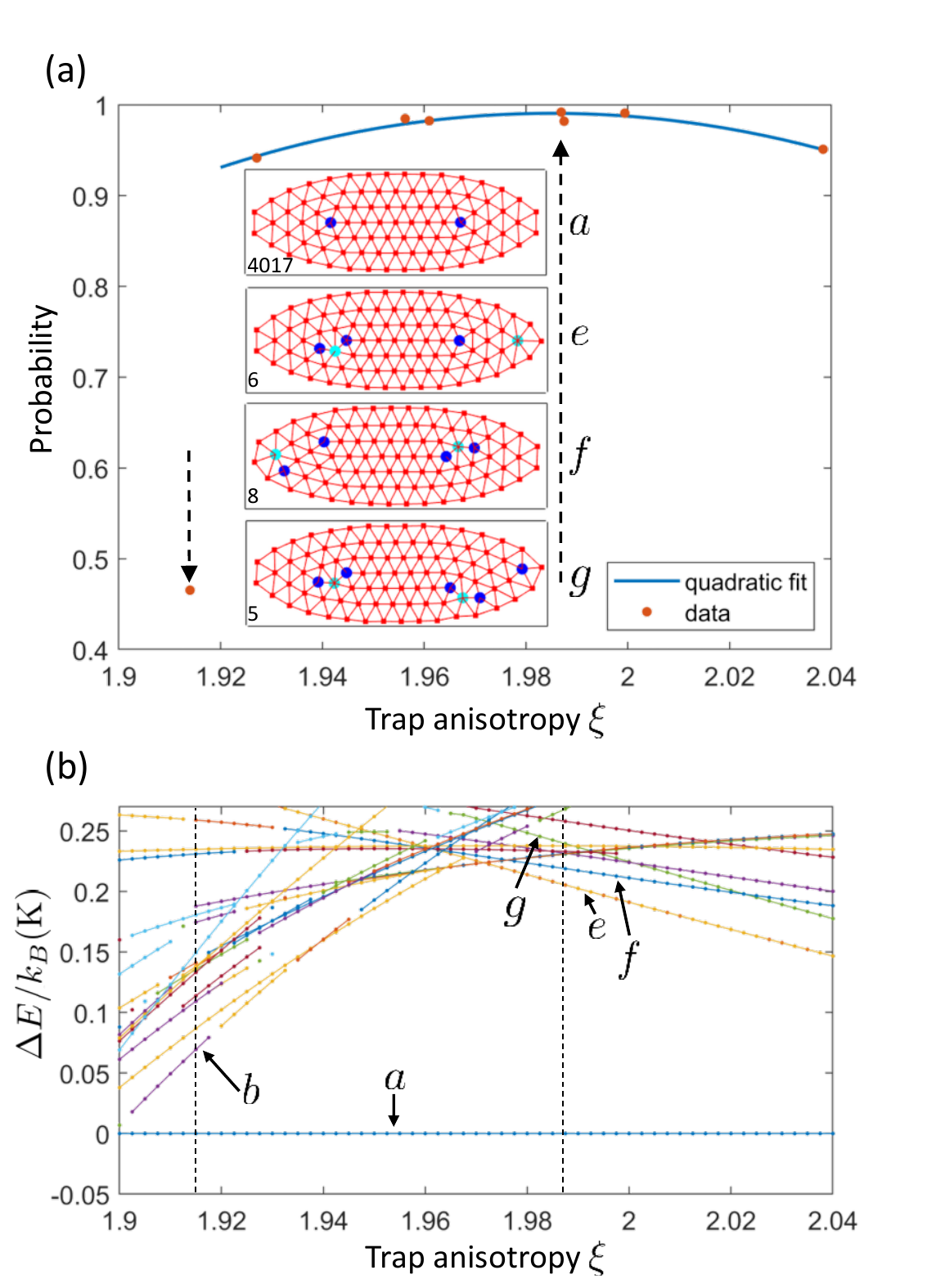}
    \caption[format=plain,justification=justified,singlelinecheck=false]{Crystal configurations versus potential anisotropy for a  91-ion crystal. (a) The red points represent experimental data of the probability of finding the main configuration, $\max(p_i)$, as a function of trap anisotropy $\xi$. The blue line, a quadratic fit, provides a guide to the eye for data in the stable regime, where the probability of being in the main configuration exceeds 90\%, and can be used to find the optimum anisotropy. After optimizing $\xi$, the main configuration (labeled $a$) is found in more than 99\% of all 4050 images taken (upward-pointing arrow), whereas configurations $e-g$ were observed only a few times as indicated by the numbers next to them. The downward-pointing arrow indicates the value of $\xi$ for which the data of Fig.~\ref{fig:cluster_algo} was taken. (b) Numerical simulations of metastable configurations reproduce many configurations found in the experiment. Moreover, they show that the energy gap between metastable configurations and the ground state becomes maximal close to the value of $\xi$ for which the crystal lattice is most stable in the experiment.}
    \label{fig:sim_annealing}
\end{figure}

To better understand how the anisotropy of the trapping potential influences the probability of observing the main configuration, we numerically searched for the ground state and low-energy metastable configurations. In these simulations, we assumed that the ions were confined in a conservative anisotropic harmonic potential characterized by oscillation frequencies $\omega_i$ and carried out an annealing routine with parameters set such that the lowest-energy configuration would be found in about 15\% of the annealing runs. For six values of trap anisotropy in the range of $1.9\le\xi\le2.06$, the annealing procedure was repeated 500 times each in order to identify a large number of metastable configurations. In a second step, the configurations found at a particular anisotropy were used as the starting point for an annealing scheme at a slightly different anisotropy using a slower annealing schedule in order to complete the picture.  

 We find that the lattice configurations observed in the experiment can be assigned to low-energy configurations obtained by simulated annealing. In Fig.~\ref{fig:sim_annealing}(b), we plot the energy gap of these configurations above the energy of the ground-state configuration versus the anisotropy $\xi$ of the potential. All configurations observed in the experiment at $\xi=1.987$ were found in the simulation, as were the most frequent configurations at $\xi=1.915$, labeled $a$ and $b$ in Fig.~\ref{fig:cluster_algo}. Moreover, the energy gap between the ground state and the first excited state is nearly at its maximum at $\xi=1.987$, where the ground-state configuration exhibits the highest stability in the experiment. Here, the energy gap corresponds to a temperature of 200~mK. In contrast, at $\xi=1.915$, the gap is only about 50~mK, which suggests that insufficient laser cooling or low-energy Langevin collisions could induce configuration changes much more easily.

\subsubsection{Mitigation of configuration changes}
The higher stability of the main configuration compared to other configurations is based on the fact that the ion crystal is symmetric with respect to both axes that span the crystal plane, which can be seen in the top image of Fig. \ref{fig:cluster_algo} (b) and (c). For given trapping parameters, only crystals consisting of a certain number of ions can fulfill this criterion, which is why we trap specifically 8, 54, 91 and 105 ions, a selection of the possibilities, for the experiments presented in this manuscript. 
After the potential anisotropy is fine-tuned in order to maximize the time spent in the main configuration, the lifetime of metastable configurations is rather short, often not exceeding 10~ms. The effect of undesired transitions into these configurations can then be mitigated in experiments, probing the desired configuration by the experiment control program: it detects the crystal configuration found in each experiment and automatically repeats those experiments where the crystal was found in the wrong one. 
Additionally, it might be feasible to deliberately destabilize an undesired metastable configuration by transiently changing the laser cooling parameters, which has not been necessary for the experiments we have conducted so far.

\subsection{Ground-state cooling of the out-of-plane motional modes}
\label{sec:groundstatecooling}

Controlling a large number of ions necessitates fast multimode cooling techniques such as EIT cooling \cite{Morigi2000} as subsequent cooling of individual modes by Raman or resolved-sideband cooling becomes time consuming and thus infeasible. EIT cooling, like sideband cooling, is only applicable to precooled ions and has been demonstrated with a single ion \cite{Roos2000}, long strings of ions \cite{Lechner2016}, 2d crystals in Penning traps \cite{Jordan2019}, and more recently a planar 12-ion crystal in an rf trap \cite{Qiao2021}. 

Here, we employ EIT cooling to prepare all $N$ out-of-plane motional modes of 2d crystals with up to 105 ions close to the ground state. The cooling scheme is implemented in a $\Lambda$-type three-level system using the Zeeman sublevels of the 4S$_{1/2}$ manifold coupled off-resonantly via the 4P$_{1/2}$ level. We use two perpendicular beams at 397\,nm, one $\pi$-polarized and one $\sigma^-$-polarized beam, each having a $k$-vector at 45$^\circ$ angle with respect to the crystal plane. The $\sigma^-$-polarized light strongly couples the  4S$_{1/2}$ ($m=+1/2$)  ground state to the excited 4P$_{1/2}$ ($m=-1/2$) state, resulting in the generation of dressed states. The weak $\pi$-polarized probe beam couples the 4S$_{1/2}$ ($m=-1/2$) to the 4P$_{1/2}$ ($m=-1/2$) state and probes the resulting Fano-like absorption spectrum. The light shift from the bare energy states is adjusted via the power of the $\sigma^-$-beam to match the center of the out-of-plane frequency spectrum, spanning a few hundred kHz. This ensures that all mode frequencies are covered by the absorption spectrum to enable efficient cooling of all modes simultaneously.

\begin{figure}[tbp]
    \centering
    \includegraphics[width=0.5\textwidth]{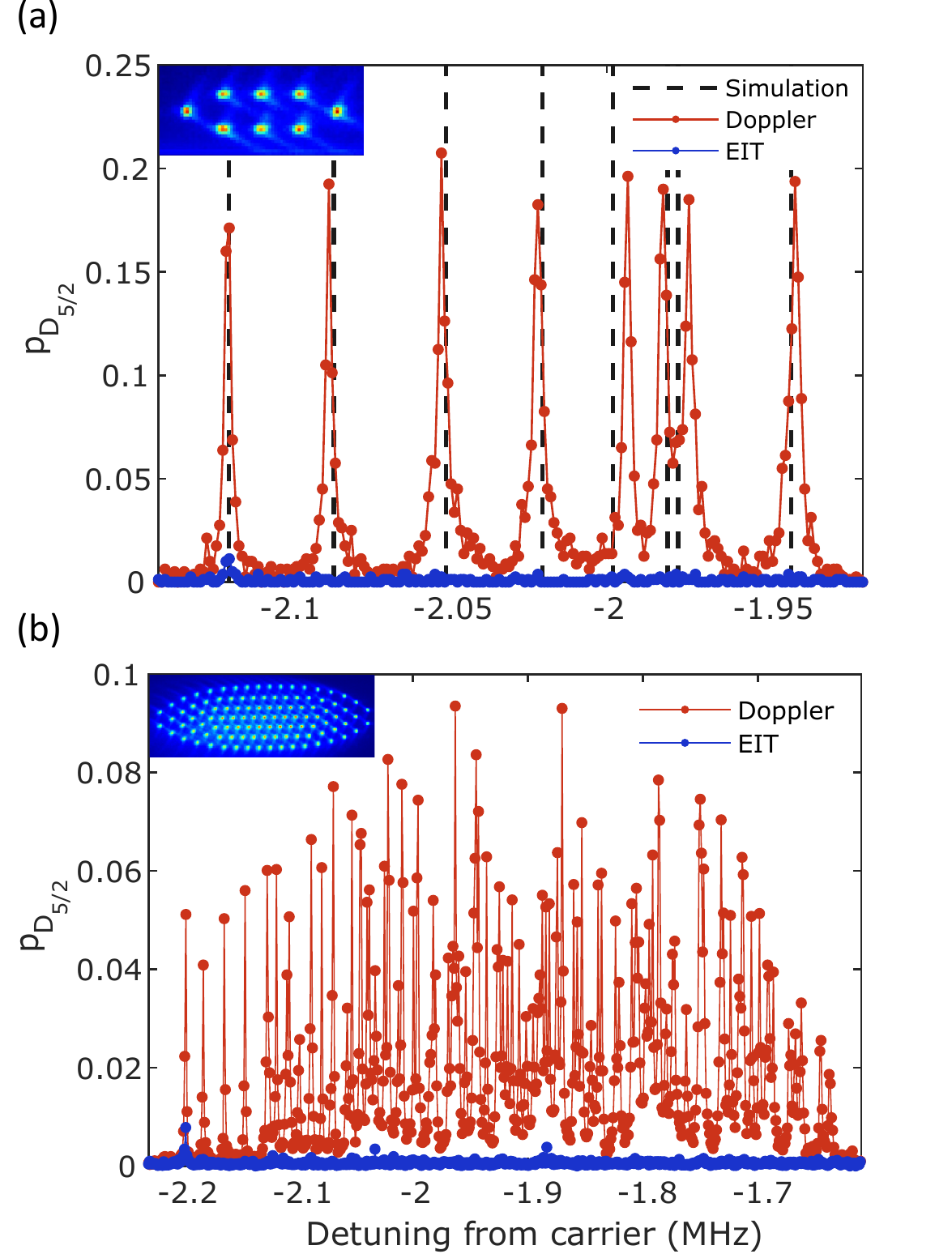}
    \caption[format=plain,justification=justified,singlelinecheck=false]{EIT cooling of (a) an 8-ion crystal and (b) a 105-ion crystal. The red data points represent the red-sideband excitation spectrum of the out-of-plane modes after only Doppler cooling and the blue data points with additional EIT cooling for 300~\textmu s. The most prominent peaks in the EIT-cooled spectra are the center-of-mass modes, which heat up during the measurement time in proportion to $\dot{\bar{n}} N$. Note that red and blue solid lines are guides to the eye. The dashed vertical lines in (a) show the simulated mode frequencies for the given trapping parameters obtained from pseudopotential theory.}
    \label{fig:eit_cooling}
\end{figure}

In order to verify that all motional modes are cooled, we perform frequency scans probing the red sideband spectrum of the 4S$_{1/2} \leftrightarrow 3$D$_{5/2}$  transition, once after only Doppler cooling, and once after additional EIT cooling for 300\,\textmu s and optical pumping for 100\,\textmu s. Figure~\ref{fig:eit_cooling} shows the measured spectra for an 8- and a 105-ion crystal, trapped in a potential with oscillation frequencies $\omega_{x,y,z} = 2\pi\times(2120, 720, 370)$\,kHz and $\omega_{x,y,z} = 2\pi\times(2188,528,248)$\,kHz, respectively.

In the Doppler-cooled spectrum we identify the red sidebands of the out-of-plane motional modes. For the 8-ion crystal, the mode frequencies as well as the mode structure represented by the Lamb--Dicke parameters of individual ions were obtained by numerical simulations based on pseudopotential theory and compared to the experimental data. For most of the modes we find good agreement between the simulated values and the measured mode frequencies as well as relative couplings to individual ions in a planar crystal. To understand the deviations for some of the modes at lower frequencies, we performed further simulations using a Floquet-Lyapunov approach \cite{Landa2012} as well as a Fourier transform of the simulated motion of the ions, both of which take the full rf field into account. However, neither of the methods improved the agreement between simulated and measured normal mode frequencies. The reason for the discrepancies is unknown up to now.

In the EIT-cooled spectra we see a clear suppression of excitation for all out-of-plane motional modes, which demonstrates an efficient ground-state cooling in a qualitative way. A quantitative investigation of such low temperatures is not straightforward, as standard methods applicable to single ions are not directly adaptable for large ion crystals. A generalized sideband thermometry technique applicable to individual modes in larger crystals has been studied and will be subject to a forthcoming publication.

Least-square fits of Rabi oscillations of the red sideband of the center-of-mass mode in an 8-ion crystal, as well as sideband thermometry of a single ion \cite{Leibfried2003}, revealed a mean phonon number of the out-of-plane center-of-mass mode of  $\bar{n}$ = 0.06 quanta after EIT cooling. Introducing a variable waiting time between EIT cooling and the temperature measurement, provided a heating rate of about $\dot{\bar{n}}$ = 15 quanta/s per ion
for the center-of-mass mode, which experiences the highest degree of heating out of all out-of-plane modes. This value is not competitive with some of the lowest heating rates reported in the literature \cite{An2019, Pogorelov2021}, but might be improved by a change of the design of the dc filter lines potentially reducing the electronic noise level on the dc electrodes.

\section{Conclusion}
\label{sec:conclusion}

In this work, we have presented an ion-trap apparatus developed towards the goal of carrying out quantum simulations for studying 2d many-body physics beyond the capabilities of classical computers.
The characterization measurements provided in this manuscript reflect the high level of control we achieve in our setup, providing a solid foundation for the planned endeavours. Fundamental requirements for the envisaged experiments are the ability to trap large ion crystals which extend in two dimensions only (to minimize adverse micromotion-induced effects on laser-ion interactions) and to ensure a stable crystal lattice during measurements. 

Our apparatus is capable of trapping stationary, 2d crystals of $\sim 100$ ions. We achieve planarity via voltage control of the trap dc electrodes and verify it by measuring low out-of-plane micromotion modulation indices.
We conclude excellent ion-crystal stability from the observed typical crystal lifetimes of several hours, 
survival in crystalline form for at least several seconds in the absence of any laser cooling, and fast, automated, recrystallization ($\sim 100$~ms) once a melting event occurs. Moreover, we have demonstrated that the influence of crystal lattice configuration changes on experiments can be practically eliminated, disarming a major challenge when working with large 2d crystals. To this end, lattice instabilities are minimized by adjustment of the trap anisotropy. In addition, data is automatically retaken for rare remaining occurrences of structural changes after in-sequence identification of the crystal configuration. A stable planar 91-ion crystal has already been employed successfully for correlation spectroscopy measurements carried out in Ref.~\cite{Hainzer2022}, demonstrating the experimental robustness of our apparatus.

Another requirement for the planned quantum simulation experiments is precise knowledge and high control of the crystals' out-of-plane motional modes, as they will be employed to mediate entangling interactions between the ions. First, anharmonicities in the trapping potential could lead to a modification of the normal-mode frequencies and amplitudes \cite{Home2011}. From the analysis of crystal images, we find that substantial contributions from anharmonicities to the trapping potential can be ruled out. Second, cooling of these modes close to the motional ground state has to be achieved. We have presented EIT cooling of a small (8-ion) and a large (105-ion) crystal, which, to our knowledge, is the largest ion crystal in an rf trap that has successfully been cooled close to the ground state. Third, we conclude sufficient stability of the motional modes, a prerequisite for creating reliable entangling interactions, from observing low drifts and a heating-rate limited motional coherence.

In order to create entangling interactions, we are currently setting up a laser system at 396 nm, to couple the ground-state qubit on a stimulated Raman transition via the excited 4P$_{1/2}$ states. By using a bichromatic light field, detuned from the red and blue sidebands of the out-of-plane motional modes of the crystal, we will be able to engineer spin-spin interactions with a tunable interaction strength \cite{Porras2004}. Moreover, we aim to implement individual-ion control via a tightly-focused 729 nm laser beam steered by a 2d acousto-optic deflector. This will enable sequential single-qubit rotations and structuring of the crystal lattice into different geometries by shelving ions into long-lived states which do not couple to the field inducing spin-spin interactions.

\section{Acknowledgements}

We thank Mark Dugan and Philippe Bado from Translume Inc., Alpine Quantum Technologies GmbH and the Innsbruck ion trapping research groups for their support. This project has received funding from the European Research Council (ERC) under the European Union’s Horizon 2020 research and innovation programme (grant agreement No 741541). Furthermore, we acknowledge funding from the European Union’s Horizon 2020 research and innovation programme under the Marie Skłodowska‐Curie grant agreement No 801110 and the Austrian Federal Ministry of Education, Science and Research (BMBWF).

\hfill

\hfill

\hfill

\appendix

\section{Comparison of different crystal orientations}
\label{sec:calculations_appendix}

We compare the creation of planar ion crystals in a linear rf trap in two different geometries: First, we consider an ion crystal that is trapped in a plane spanned by the two radial trap directions. Second, we consider a crystal in the plane spanned by one radial and the axial trap direction. We compare these two cases in terms of rf power requirements as well as stability of the trap frequencies in the weakly confined directions in the presence of rf power fluctuations. Lower stability can lead to more crystal lattice configuration changes, which should be suppressed as much as possible.

We assume that one wants to trap the crystal in a potential characterized by the three secular frequencies $\omega_i$. Moreover, the $q$-parameter should be upper-bounded by a value $q_m$ to keep the micromotion amplitude low. The following calculation is done within the pseudopotential approximation.

In a linear trap, the oscillation frequencies can be expressed in terms of the $q$- and $a$-parameters and the trap drive frequency $\Omega$,
\begin{align}
\omega_{1,2} &= \frac{\Omega}{2}\sqrt{\frac{q^2}{2}+a_{1,2}}\\
\omega_3 &= \frac{\Omega}{2}\sqrt{a_3},
\end{align}
with $a_1+a_2+a_3=0$, where $a_i$ are set by static voltages. Because of
$\sum_n\omega_n^2=q^2\Omega^2/4$
this lower-bounds the value of the angular drive frequency $\Omega$ to
\begin{equation}
\Omega^2=\frac{4}{q^2}\sum_{n=1}^3\omega_n^2\ge\frac{4}{q_m^2}\sum_{n=1}^3\omega_n^2.\label{eq:OmegaRF}
\end{equation}
As Eq.~(\ref{eq:OmegaRF}) does not depend on the orientation of the planar crystal in the trap, the minimum rf power required to generate the desired set of secular frequencies is the same for both geometries.

Secondly, we investigate the influence of trap-voltage fluctuations on the secular frequencies. As it is much harder to generate stable rf voltages than dc voltages, we will focus on the influence of rf-power fluctuations, which will change the $q$-parameter. The secular frequencies depend on $q$ via
\begin{align}
\frac{d\omega_i}{dq}&=\frac{\Omega}{4}\frac{q}{\sqrt{\frac{q^2}{2}+a_i}}
=\frac{\Omega^2q}{8\omega_i}
=\frac{1}{2q\omega_i}\sum_{n=1}^3\omega_n^2\;\;,
i\in{1,2}
\label{eq:powerdependence}\\
\frac{d\omega_3}{dq}&=0.
\end{align}
For the realization of spin-spin interactions, the lasers are typically detuned by a few tens of kHz from the center-of-mass mode with the highest frequency. In view of suppressing the coupling strength variations caused by fluctuating rf voltages, the best option is to confine the crystal in the radial plane of the linear trap. However, Eq.~(\ref{eq:powerdependence}) shows that for a crystal, whose plane includes the axial direction, the effect of fluctuating trap voltages on the detuning of the bichromatic beam is about two times weaker than for the case of an ion string in a linear trap with the same maximum secular frequency.
A second point to be considered are changes in the aspect ratio of a planar crystal induced by fluctuating rf voltages $V_{\textrm{rf}}$. For a planar crystal of circular shape trapped in the radial plane, the aspect ratio does not change. For a crystal aligned with the axial direction, the oscillation frequency of the weak axis that is pointing into the radial direction is affected, leading to changes of the trap anisotropy $\xi$. An rf-voltage change $\Delta V_{\textrm{rf}}$ will change $\xi$ by
\[
\frac{\Delta\xi}{\xi}=-\left(\frac{\Delta V_{\textrm{rf}}}{V_{\textrm{rf}}}\right)\frac{\sum_n\omega_n^2}{2\omega_2^2},
\]
if $\omega_1\gg\omega_2\ge\omega_3$ . This shows that $\xi$ is more sensitive to changes of $V_{\textrm{rf}}$ than $\omega_1$. Therefore, care should be taken to actively stabilize the rf voltage.

\section{Technical details of the ion trap}
\label{sec:trap_appendix}

\subsection{Ion trap geometry}
 
We optimize the electrode structure of the ion trap using finite-element simulations (\textit{COMSOL Multiphysics}), to best fulfill the constraints imposed by the envisaged quantum simulation experiments. Here, in particular, it is important to create sufficiently fast interactions, which rely on high secular frequencies. We aim for a secular frequency of 2\,MHz in the strongly confined direction. Additionally, it is beneficial to have $q<0.1$ as the micromotion amplitude scales linearly with $q$. The $q$-parameter and secular frequencies relate to the trap drive frequency, the rf voltage that is applied to the trap, and the distance between the ions and the trap electrodes. There exists a limit to how much voltage the trap chip can reasonably handle, which we estimated to be on the order of 1\,kV peak-to-peak. The aforementioned constraints for the desired secular frequencies, $q$-parameter, and maximum rf drive voltage already put a bound on the rf drive frequency, which can be chosen by design of the helical resonator circuit that is employed to amplify the trap rf voltage. In order to reduce anomalous heating of the ions \cite{Brownnutt2015}, it is beneficial to keep the distance between ion and trap electrode as large as possible. From simulations we find a tradeoff between all parameters and end up with a distance of $d=400$~\textmu m between the rf electrode and trap center.

The length of the central dc electrode segments of the trap is chosen based on the ability to perform tilting operations on the ion crystals with respect to the principal axes. A length of 1.5\,mm is short enough to allow for efficient tilting via reasonably low asymmetric voltages on the outer dc segments.

\subsection{Control and stability of the trap rf}
\label{sec:stab_appendix}

In order to produce the high-voltage rf field required to drive the trap, we amplify the signal from a frequency generator with an rf amplifier as well as a helical resonator \cite{Siverns2012}. The resonance frequency depends on the circuit's inductance, dominated by its coil (40.4\,mm diameter, 5\,mm wire thickness, 10\,mm winding pitch, and 70\,mm coil height), and capacitance, dominated by the trap itself. We find a resonance at 43.22\,MHz from which we estimate the upper limit for the capacitance of the trap to be about 12\,pF. 

Without active stabilization we observe typical drifts of the radial frequencies of several kHz over a timescale of minutes, caused by temperature fluctuations around the experiment, which need to be corrected for. A home-built circuit based on a proportional-integral controller is used to stabilize the rf power. The input signal for the  circuit is detected inductively with a pickup coil close to the coil of the helical resonator. Spectroscopic measurements on a single ion reveal that for a secular frequency of $\omega_x=2\pi\times 2.25$~MHz the peak-to-peak variation of the oscillation frequency are below 100~Hz over the course of one hour, with the main variation being given by a linear drift. Moreover, Ramsey experiments on motional superposition states $|0\rangle+|1\rangle$ yield a coherence time of about 40~ms that is limited by motional heating. The combination of these two measurements demonstrates that the rf-stabilization provides excellent frequency stability on both long and short time scales. 

\subsection{Trapping potential shaping and micromotion compensation}

The shape of the trapping potential can be precisely controlled via the 12 segmented dc electrodes of the ion trap. Low-pass filters (first order) with cutoff frequencies of about 300\,Hz are used to suppress noise in the secular frequency regime while allowing for the possibility of fast dc voltage control on millisecond time scales.

In order to execute certain actions on the ion crystals, such as micromotion compensation (shifts in $x$, $y$ or $z$), rotation of the potential (about the $x$, $y$ or $z$ axis) or adjustment of the confinement along one of the three principle axes, we calculate the required voltages that are applied to the 12 dc electrodes to implement these actions. A fit to the trap potential of the individual electrodes obtained from finite element simulations, done in \textit{COMSOL Multiphysics} based on our trap design, forms the basis for these calculations. The trap potential of each electrode is expanded into spherical harmonic terms. In this way, the various actions on the ion crystal can be decoupled from each other. In a second step, an overdetermined inversion problem has to be solved linking the required action on the ions to the voltages that need to be applied on all 12 dc electrodes as described in Appendix B of Ref.~\cite{vanMourik2020}. We use the Tikhonov method to solve the equations based on a singular value decomposition \cite{Singer2010}.


%

\end{document}